\documentclass[aps,epsfig,rotate,showpacs]{revtex4}  
\usepackage{epsfig}
\usepackage{graphicx}
\usepackage{amsmath}

\begin{document}                
\title{
Intrinsic Decoherence Dynamics in Smooth Hamiltonian Systems: Quantum-classical
Correspondence}
\author{Jiangbin Gong and Paul Brumer}
\affiliation{Chemical Physics Theory Group,\\Department of Chemistry,\\
University of Toronto\\ Toronto, Canada  M5S 3H6}
\date{\today}

\begin{abstract}              
A direct classical analog of
the quantum dynamics of intrinsic decoherence in Hamiltonian systems,
characterized by the time dependence of the linear entropy of the reduced density operator,
is introduced.
The similarities and differences between 
the classical and quantum decoherence dynamics of an initial quantum state
are exposed using both analytical and computational results.
In particular, the classicality of early-time intrinsic decoherence dynamics
is explored analytically using a second-order perturbative treatment, and an interesting
connection between
decoherence rates and the stability nature
of classical trajectories is revealed in a simple approximate 
classical theory of intrinsic decoherence dynamics.
The results offer new insights into decoherence, dynamics of quantum
entanglement,
and quantum chaos.
\end{abstract}

\maketitle

\section{Introduction}
Quantum dynamics induces unitary transformations in Hilbert space, but most often
it is only the dynamics projected onto a Hilbert {\it subspace} 
that is of interest. In general this reduced dynamics is nonunitary and
therefore displays decoherence \cite{zurekreview}.
For example, if a system of interest is coupled to a bath, then
averaging over the bath degrees of freedom introduces decoherence in the system dynamics. 
Likewise, in an isolated system, the reduced dynamics of
a subsystem of this isolated system can display decoherence. 
We have termed decoherence in the latter case
``{\it intrinsic decoherence}'' since it does not involve an external bath \cite{batista}. 

Understanding decoherence is of crucial importance
to a variety of modern fields such as quantum information
processing \cite{qcbook} and quantum control of atomic and molecular
processes \cite{brumerreview,ricebook,brumerbook}.  Our interest here is in
the quantum-classical correspondence (QCC) between classical and 
quantum descriptions of the {\it dynamics} of decoherence.
Specifically, we consider an initial quantum state subjected to either quantum
or classical dynamics and compare the time evolution of the decoherence
in both cases.  We note that
the formal theory of correspondence between quantum dynamics and
classical Liouville dynamics \cite{wilkie}  suggests that 
classical Liouville dynamics projected
onto a subspace should also display decoherence. 
That is, as in the quantum
case, the classical Liouville dynamics considered in the entire phase space is unitary and
the classical Liouville dynamics projected onto a subspace 
is nonunitary.  We therefore expect that
the reduced classical Liouville dynamics propagated
classically will show decoherence dynamics that is, at least qualitatively,
parallel to that seen in the reduced quantum dynamics insofar as the loss of phase
information, entropy production, etc.
In the case of
bath-induced decoherence we recently showed analytically that (a)
one can indeed introduce a direct classical analog of quantum
decoherence, and (b) examining the dynamics of decoherence
classically gives new insights into both the dynamics of
decoherence described quantum mechanically and into the conditions
for QCC in decoherence dynamics \cite{gongprl}. 

Here we extend these considerations to intrinsic decoherence, both analytically and computationally.
Specifically,  in this paper we study  QCC in the dynamics 
of intrinsic decoherence in smooth Hamiltonian systems, with an emphasis on the usefulness
of classical dynamics in describing intrinsic decoherence.
In particular, the classicality of early-time intrinsic decoherence dynamics
is studied using a second-order perturbative treatment, and the interesting
connection between
decoherence rates at later times and the stability
properties of classical trajectories is revealed by considering
a simple approximate 
classical theory of intrinsic decoherence dynamics.
The analytic and computational results shed new light on 
decoherence, dynamics of quantum entanglement,
and quantum chaos.  This study is also of interest to semiclassical 
decoherence studies \cite{bmiller}, e.g., semiclassical descriptions of
intrinsic decoherence dynamics
in large molecular systems \cite{batista}.

This paper is organized as follows. In Sec. \ref{perturbation-section}
we introduce a second-order perturbation theory in an effort to understand
QCC in early-time intrinsic decoherence dynamics.  
For simplicity we focus upon two degree-of-freedom systems, but the extension to larger systems
is straightforward.
Computational results of two sample cases in coupled-oscillator model systems,
which strongly support
the physical picture afforded by the perturbative treatment, are presented in
Sec. \ref{sample}.
Then,  a classical theory
of intrinsic decoherence dynamics for initially localized states is derived in Sec. 
\ref{classical-theory}.  In the same section, 
detailed  computational studies using this
simple theory are carried out for the
quartic oscillator model and one of its variants.
Discussions and a summary comprise Sec. \ref{summary}.

\section{Early-time Intrinsic Decoherence Dynamics}
\label{perturbation-section}
Consider a conservative
system composed of two subsystems, with the total Hamiltonian given by 
\begin{eqnarray}
H(Q,P,q,p)=\frac{P^{2}}{2}+\frac{p^{2}}{2}+V_{1}(Q)+V_{2}(q)+V_{12}(Q,q),
\end{eqnarray}
where $(Q,P)$ and ($q,p$) are dimensionless
phase space conjugate variables,
$V_{i}$ is the potential of the $i$-th subsystem, and
$V_{12}(Q,q)$ describes arbitrary coupling between the two subsystems.
As the system evolves
the total system wavefunction $|\psi(t)\rangle$ 
becomes inseparable due to quantum entanglement,
even if it is initially separable in $Q$ and $q$.
As a result, measuring a subsystem would collapse the system wavefunction
and therefore affect the properties 
of the other subsystem. Similarly,
ignoring a subsystem  decoheres the other one.
The degree of intrinsic decoherence, 
which is induced by, and is a manifestation of, quantum entanglement between the two subsystems,
can be measured by 
a well-known quantity: the quantum linear entropy \cite{linear-entropy}
\begin{eqnarray}
S_{{\rm q}}=
1-Tr_{1}(\hat{\tilde{\rho}}^{2}),
\end{eqnarray}
where $Tr_{i}$  denotes a trace over the $i$-th subsystem, and 
$\hat{\tilde{\rho}}\equiv Tr_{2} \left(|\psi(t)\rangle\langle\psi(t)|\right)$
is the reduced density operator
for the first subsystem. 
An increase in $S_{{\rm q}}$ suggests
an increase of $1/(1-S_{{\rm q}})$, which gives
the number of orthogonal quantum states that are incoherently populated if the second subsystem
is ignored. Below we choose $q, p$ as the ``bath" variables and $P,Q$ as the system variables.

Let $\rho_{c}(Q,P,q,p,t)$ denote the phase space
distribution function evolved classically, and $\rho_{W}(Q,P,q,p,t)$ denote the
quantum (Wigner) phase space
distribution function.
Their time evolution equations are given by
\begin{eqnarray}
\frac{\partial \rho_{c}}{\partial t}& = &\{H,\rho_{c}\}, \\
\frac{\partial \rho_{W}}{\partial t}&=& \{H,\rho_{W}\}_{M},
\end{eqnarray}
where $\{\cdot\}$ denotes classical Poisson bracket and
$\{\cdot\}_{M}$ denotes quantum Moyal bracket \cite{moyal}.
We define classical and quantum reduced distribution functions as
\begin{eqnarray} \tilde{\rho}_{c}(Q,P,t) &\equiv &
\int \rho_{c}(Q,P,q,p,t)\ dq \ dp, \\
\tilde{\rho}_{W}(Q,P,t) & \equiv & \int \rho_{W}(Q,P,q,p,t)\ dq\ dp.
\end{eqnarray}
Since \begin{eqnarray}
S_{{\rm q}}(t)=1-2\pi\hbar\int\tilde{\rho}_{W}^{2}(Q,P,t)\ dQ\ dP, 
\end{eqnarray}where $\hbar$ is the effective
Planck constant,
we can define a
classical analog [denoted $S_{c}(t)$] to $S_{{\rm q}}(t)$ 
by replacing $\tilde{\rho}_{W}$ with 
$\tilde{\rho}_{c}$. That is,
\begin{eqnarray}
S_{c}(t)\equiv 1-2\pi\hbar\int\tilde{\rho}_{c}^{2}(Q,P,t)\ dQ\ dP.
\end{eqnarray}
The main focus here is to
compare $S_{c}(t)$ 
with $S_{{\rm q}}(t)$, i.e., the classical
vs. quantum evolution of the intrinsic decoherence dynamics, as measured by the
classical vs. quantum entropy.

Perturbative treatments have proved to very useful in understanding
decoherence dynamics \cite{gongprl,kim,duan,bacon}. Here, to analytically examine
classical vs. quantum intrinsic decoherence dynamics at early times,
we apply the perturbative approach developed
in our previous work \cite{gongprl} to the case of intrinsic decoherence dynamics.
Specifically, consider a  second-order
perturbative expansion with respect to the time variable $t$ for both $S_{{\rm q}}$ and $S_{c}$, i.e.,
\begin{eqnarray}
S_{c}(t)& = & S_{c}(0)+\frac{t}{\tau_{c,1}}+\frac{t^{2}}{\tau_{c,2}^{2}}+\cdots, \nonumber \\
S_{{\rm q}}(t)& = & S_{{\rm q}}(0)+\frac{t}{\tau_{{\rm q},1}}+\frac{t^{2}}{\tau_{{\rm q},2}^{2}}+\cdots.
\end{eqnarray}
Then, from the classical and quantum dynamics of the entire system
one obtains
\begin{eqnarray}
\frac{1}{\tau_{c,1}}=-4\pi\hbar\int \tilde{\rho}_{c}(Q,P,0)
\int \{H, \rho_{c}(Q,P,q,p,0)\}\ dq\ dp \ dQ\ dP,
\end{eqnarray}
\begin{eqnarray}
\frac{1}{\tau_{{\rm q},1}}=-4\pi\hbar\int  \tilde{\rho}_{W}(Q,P,0) 
\int \{H, \rho_{W}(Q,P,q,p,0)\}_{M}\ dq\ dp \ dQ \ dP,
\end{eqnarray}
\begin{eqnarray}
\frac{1}{\tau_{c,2}^{2}}& = &-2\pi\hbar\int \tilde{\rho}_{c}(Q,P,0)
\int \left\{H, \left\{H,
\rho_{c}(Q,P,q,p,0)\right\}\right\}\ dq\ dp \ dQ \ dP
\nonumber \\
& & -2\pi\hbar\int   \left[ \int \left\{H,
\rho_{c}(Q,P,q,p,0)\right\}\ dq\ dp\right]^{2} \ dQ\ dP,
\label{t2c-a}
\end{eqnarray}
and
\begin{eqnarray}
\frac{1}{\tau_{{\rm q},2}^{2}}& = &-2\pi\hbar\int \tilde{\rho}_{W}(Q,P,0) 
\int \left\{H, \left\{H,
\rho_{W}(Q,P,q,p,0)\right\}_{M}\right\}_{M}\ dq \ dp  \ dQ \ dP
\nonumber \\
& & -2\pi\hbar\int   \left[ \int \left\{H,
\rho_{W}(Q,P,q,p,0)\right\}_{M}\ dq\ dp \right]^{2}\ dQ\ dP.
\label{t2q-a}
\end{eqnarray}
Further, using the definitions of the classical Poisson and quantum Moyal brackets and
assuming that initial classical and quantum distribution functions are identical and
separable, i.e., 
\begin{eqnarray}
\rho_{c}(Q,P,q,p,0)=\rho_{W}(Q,P,q,p,0)=\tilde{\rho}_{1}^{0}(Q,P)\tilde{\rho}_{2}^{0}(q,p),
\end{eqnarray}
we have
\begin{equation}
\frac{1}{\tau_{c,1}}=\frac{1}{\tau_{{\rm q},1}}=0,
\label{firstorder}
\end{equation}
\begin{equation}
\frac{1}{\tau_{c,2}^{2}}=2\pi\hbar \int 
\left[\frac{\partial\tilde{\rho}_{1}^{0}(Q,P)}{\partial
P}\right]^{2}C(0,0)\ dQ \ dP,
\label{cla-2nd}
\end{equation}
and
\begin{eqnarray}
\frac{1}{\tau_{{\rm q},2}^{2}}&=&\frac{1}{\tau_{c,2}^{2}} 
+2\pi\hbar\int \sum_{l_{1}\ne l_{2}\ge
0}\frac{[\hbar/(2i)]^{(2l_{1}+2l_{2})}}{(2l_{1}+1)!(2l_{2}+1)!} \nonumber \\
&&\times
\frac{\partial^{(2l_{1}+1)}\tilde{\rho}_{1}^{0}(Q,P)}{\partial P^{(2l_{1}+1)}}
\frac{\partial^{(2l_{2}+1)}\tilde{\rho}_{1}^{0}(Q,P)}{\partial P^{(2l_{2}+1)}}C(l_{1},l_{2})\ dQ\ dP,
\label{quan-2nd}
\end{eqnarray}
where $C(l_{1},l_{2})$ is a correlation function given by
\begin{eqnarray}
C(l_{1},l_{2})&\equiv&
\left\langle \frac{\partial^{(2l_{1}+1)} V(Q,q)}{\partial Q^{(2l_{1}+1)}}
\frac{\partial^{(2l_{2}+1)} V(Q,q)}{\partial Q^{(2l_{2}+1)}}
\right\rangle_{\tilde{\rho}_{2}^{0}}\nonumber \\
&-& 
\left\langle\frac{\partial^{(2l_{1}+1)} V(Q,q)}{\partial Q^{(2l_{1}+1)}}\right\rangle_{\tilde{\rho}_{2}^{0}}
\left\langle\frac{\partial^{(2l_{2}+1)} V(Q,q)}{\partial
Q^{(2l_{2}+1)}}\right\rangle_{\tilde{\rho}^{0}_{2}}.
\end{eqnarray}
Here $\langle\cdot\rangle_{\tilde{\rho}^{0}_{2}}$ denotes the ensemble average over
the zero-time ``bath distribution function"  $\tilde{\rho}_{2}^{0}(q,p)$. 
It is worth emphasizing that in our derivations we have used the same
initial state for the  classical and quantum dynamics.

Equation (\ref{firstorder}) shows that
the zero first-order linear entropy increase rate, i.e.,
$1/\tau_{{\rm q},1}$=0,
has a strict classical analog. Further, Eq. (\ref{cla-2nd})
indicates that  
classical Liouville dynamics also predicts a second-order entropy production
rate $1/\tau_{c,2}^{2}$
that is the analog of the second-order quantum decoherence rate $1/\tau_{{\rm q},2}^{2}$.
Thus, we can identify two categories of early-time intrinsic decoherence dynamics:
{\it classical} if $\tau_{c,2}\approx \tau_{{\rm q},2}$,
and {\it nonclassical} if $\tau_{{\rm q},2}$ appreciably differs from $\tau_{c,2}$.

To simplify Eqs. (\ref{cla-2nd}) and (\ref{quan-2nd}) we introduce the Fourier transform [denoted $
F(Q_{1},Q_{2})$]
of $\tilde{\rho}_{1}^{0}(Q,P)$, i.e.,
\begin{eqnarray}
F(Q_{1},Q_{2})&\equiv&
\int  \tilde{\rho}_{1}^{0}(\overline{Q}
,P)\exp\left[\frac{i\Delta QP}{\hbar}\right]\ dP,
\end{eqnarray}
where $\overline{Q}\equiv (Q_{1}+Q_{2})/2$ and $\Delta Q=Q_{1}-Q_{2}$.
We then obtain
\begin{equation}
\frac{1}{\tau_{c,2}^{2}}=\frac{1}{\hbar^{2}} \int
|F(Q_{1},Q_{2})|^{2} \Delta Q^{2} C(0,0)\ dQ_{1}\ dQ_{2},
\label{cla-2nd-n}
\end{equation}
and
\begin{eqnarray}
\frac{1}{\tau_{{\rm q},2}^{2}}=\frac{1}{\tau_{c,2}^{2}}
+\frac{1}{\hbar^{2}}\int |F(Q_{1},Q_{2})|^{2}\sum_{l_{1}\ne l_{2}\ge
0}\frac{\Delta Q^{(2l_{1}+2l_{2}+2)}}{(2l_{1}+1)!(2l_{2}+1)!}\frac{1}{2^{(2l_{1}+2l_{2})}}
C(l_{1},l_{2})\ dQ_{1}\ dQ_{2}.
\label{quan-2nd-n}
\end{eqnarray}
Equations (\ref{cla-2nd-n}) and (\ref{quan-2nd-n}) are general results.
For the special case of $V_{12}(Q,q)=f(Q)g(q)$,
 Eqs. (\ref{cla-2nd-n}) and
(\ref{quan-2nd-n}) can be rewritten in a simple and more enlightening form:
\begin{eqnarray}
\frac{1}{\tau_{c,2}^{2}}=
\frac{\langle g^{2}(q)\rangle_{\tilde{\rho}_{2}^{0}} -
\langle g(q)\rangle^{2}_{\tilde{\rho}_{2}^{0}}}{\hbar^{2}}\int|F(Q_{1},Q_{2})|^{2}
\Delta Q^{2}\left[\frac{d f(\overline{Q})}{dQ}\right]^{2}\ dQ_{1}\ dQ_{2}, 
\label{cfg}
\end{eqnarray}
and
\begin{eqnarray}
\frac{1}{\tau_{{\rm q},2}^{2}}=
\frac{
\langle g^{2}(q) \rangle_{\tilde{\rho}_{2}^{0}} -
\langle g(q)\rangle^{2}_{\tilde{\rho}_{2}^{0}}}{\hbar^{2}}\int
|F(Q_{1},Q_{2})|^{2}
\Delta Q^{2} \left[\frac{\Delta f(\overline{Q})}{\Delta Q}\right]^{2}\ dQ_{1}\ dQ_{2},
\label{qfg}
\end{eqnarray}
where $\Delta f(\overline{Q})/\Delta Q$ is the finite-difference function
\begin{eqnarray}
\frac{\Delta f(\overline{Q})}{\Delta Q}&\equiv &
\frac{f(\overline{Q}+\Delta Q/2)-f(\overline{Q}-\Delta Q/2)}{\Delta Q}
\nonumber \\
&=&\frac{f(Q_{1})-f(Q_{2})}{Q_{1}-Q_{2}}.
\end{eqnarray}
As a result: (1) If $f(Q)$ depends only
linearly or quadratically upon the coupling coordinate $Q$, a common approximation, then
$(1/\tau_{{\rm q},2}^{2}-1/\tau_{c,2}^{2})=0$ for any initial state. That is,
in this case there exists
perfect QCC in early-time dynamics of intrinsic decoherence, regardless of $\hbar$,
and irrespective of the potentials $V_{1}(Q)$ and $V_{2}(q)$.
(2) Even in the case of highly nonlinear $f(Q)$, as long as 
$F(Q_{1},Q_{2})$  decays fast enough
with $|Q_{1}-Q_{2}|$
such
that $\Delta f/\Delta Q \approx df/dQ$, QCC would still be excellent.
The smaller the $\hbar$, the more rigorous
is this requirement. (3) If $\Delta f/\Delta Q$ differs significantly from
$ df/dQ$ over the $Q$-coordinate  scale of the initial state, 
quantum entropy production can be totally unrelated to classical entropy production.
Such cases of poor QCC are of fundamental interest, but are not the focus of this paper.

The second-order perturbative treatment is most reliable for early-time
dynamics and for relatively weak decoherence.
The above results are particularly
significant for studies on the control of intrinsic decoherence,
where early-time
dynamics of weak decoherence is important.  
In these circumstances
it is useful to understand the extent to which (quantum) intrinsic
decoherence is equivalent to classical entropy production, i.e. to
increasing $S_c(t)$. In particular, if there exists good
correspondence between classical and quantum decoherence dynamics,
then the essence of decoherence control is equivalent to the
suppression of classical entropy production, and various classical
tools may be considered to achieve decoherence control. If not,
then fully quantum tools are required.   

The above perturbation
results clearly demonstrate that quantum dynamics of intrinsic decoherence 
has a direct analog in classical Liouville dynamics. 
This rather intriguing result motivates us to
computationally examine QCC in the dynamics of intrinsic decoherence
over all time scales. 

\section{Computational Results: Two Sample Cases}
\label{sample}

To computationally examine QCC in the dynamics of
intrinsic decoherence, we consider
coupled-oscillator model systems with smooth Hamiltonians.
In all the model systems studied below, we choose
\begin{eqnarray}
V_{1}(Q)+V_{2}(q)
=\frac{\beta}{4}(Q^{4}+q^{4}),
\end{eqnarray}
where $\beta=0.01$.
Since $V_{1}(Q)$ and $V_{2}(q)$ have no simple harmonic terms, any
observed agreement between
classical and quantum behavior  cannot be attributed to the similarity between
classical and quantum harmonic oscillator dynamics.
If the coupling potential $V_{12}(Q,q)$ is quadratic in both $Q$ and $q$, i.e.,
$V_{12}(Q,q)=\alpha Q^{2}q^{2}/2$, then
the resultant coupled-oscillator system is the well-known
quartic oscillator model \cite{eckhardt,gongpre,channel}. 
Because this model is well-studied and can display
strongly chaotic (e.g.,
$\alpha=1.0$, $\beta=0.01$) or integrable (e.g.,
$\alpha=0.03$, $\beta=0.01$) dynamics,  it is used in Sec. 
\ref{classical-theory}
as an ideal model to
study QCC in intrinsic decoherence dynamics for
both integrable and chaotic cases. 

Our perturbation theory approach predicts good classical-quantum agreement
at short times for some potentials and initial conditions and poor
agreement for others. We examine both these cases computationally.

\begin{figure}[ht]
\begin{center}
\epsfig{file=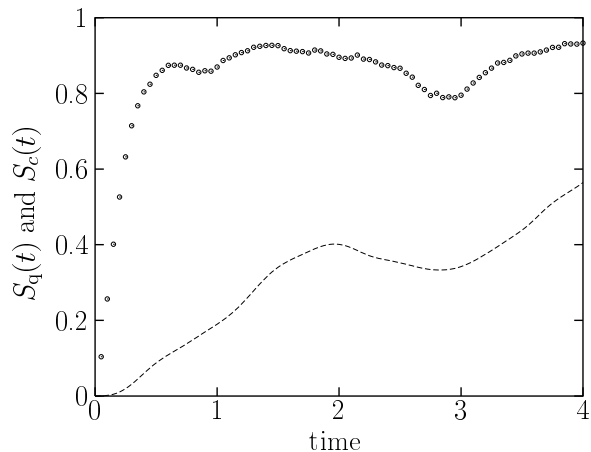,width=9.0cm}
\end{center}
\caption{A comparison between $S_{{\rm q}}(t)$ (dashed line) and $S_{c}(t)$ (discrete
circular points) in the first
sample case. The coupling potential is highly nonlinear such that at early times
classical entropy production is much faster than quantum entropy production.
See the text for details.
All variables
are in dimensionless units.}
\label{nonlinear1}
\end{figure}

It suffices to consider one case of poor agreement, since poor QCC at early times
invariably translates to similar behavior at later times.
Consider then  $V_{12}(Q,q)$ to
be some highly nonlinear potential. 
Computations of the quantum dynamics and thus the time dependence of $S_{{\rm q}}(t)$
are straightforward \cite{gongpre}. $S_{c}(t)$ is computed directly
using 
Monte-Carlo simulations
with an importance sampling
technique (where the Monte-Carlo simulations are based upon Eq. (\ref{seq1}) below).
From the analytical results above we see that  $V_{12}(Q,q)$
and the scale of the initial state play decisive roles in QCC in early-time
intrinsic decoherence dynamics.  In particular,  we expect poor
QCC if $V_{12}(Q,q)=f(Q)g(q)$ differs significantly from a linear or quadratic function of $Q$ such that
$\Delta f/\Delta Q$ differs significantly from
$(df/dQ)$ over the $Q$-coordinate scale (i.e., the support)  of the initial state. 
To confirm this computationally 
we consider $f(Q)g(q)=\sin^{2}(10Q)q^{2}$, with the initial distribution functions of the two subsystems
given by
\begin{eqnarray}
\tilde{\rho}_{1}^{0}(Q,P)&=&\frac{1}{\pi\hbar}
\exp\left[-\frac{(Q-Q_{0})^{2}}{2\sigma_{Q}^{2}}-\frac{(P-P_{0})^{2}}{2\sigma_{P}^{2}}\right],
\nonumber
\\
\tilde{\rho}_{2}^{0}(q,p)&=&\frac{1}{\pi\hbar}\exp\left[-\frac{(q-q_{0})^{2}}{2\sigma_{q}^{2}}-
\frac{(p-p_{0})^{2}}{2\sigma_{p}^{2}}
\right].
\label{initialstate}
\end{eqnarray}
Here the dimensionless effective Planck constant is chosen to be
$\hbar=0.005$ throughout, except for one case in Sec.
\ref{summary},
and $\sigma_{Q}/25=25\cdot \sigma_{P}=\sqrt{\hbar/2}$, $\sigma_{q}=\sigma_{p}=\sqrt{\hbar/2}$,
$Q_{0}=0.5$, $P_{0}=0.5$, $q_{0}=0$, with $H(Q_{0}, P_{0}, q_{0}, p_{0})=0.24$.
Note that $\tilde{\rho}_{1}^{0}(Q,P)$ is strongly squeezed in $P$ and that this initial distribution
function is considerably delocalized in $Q$.
Further, since  $|f(Q)|=|\sin^{2}(10Q)|\leq 1.0$, 
$|\Delta Q|\cdot |df/dQ|$ can be much larger
than $|\Delta f(Q)|$. Thus, for this case 
the perturbation result predicts that at early times
there can be substantial classical entropy production with insignificant 
quantum decoherence.  As shown in Fig. \ref{nonlinear1},
this is nicely confirmed by the numerical results of
$S_{{\rm q}}(t)$ and $S_{c}(t)$.
In particular, Fig. \ref{nonlinear1} shows that, at $t=1.0$
$S_{c}(t)$ (discrete points) is $\sim$ 0.9 while $S_{{\rm q}}(t)$ (solid line) is still less 
than 0.2.
Evidently,  QCC in this case 
is indeed very poor from the very beginning.

\begin{figure}[ht]
\begin{center}
\epsfig{file=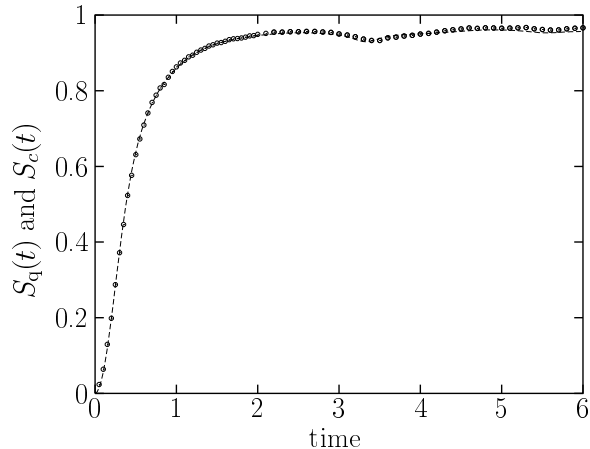,width=9.0cm}
\end{center}
\caption{ A comparison between $S_{{\rm q}}(t)$ (dashed line) and $S_{c}(t)$ (discrete
circular points) in the second
sample case.   The coupling potential is highly nonlinear in terms of the position of the second
subsystem,
but is quadratic in terms of the position of the first subsystem, resulting
in excellent quantum-classical correspondence in intrinsic decoherence
dynamics even though the initial distribution function
of the first subsystem is considerably delocalized.
See the text for details. All variables
are in dimensionless units.}
\label{nonlinear2}
\end{figure}

There remains then the important question of the quantitative degree of QCC in circumstances where our
perturbative analysis predicts good short-time QCC.   In particular, 
it is important to investigate whether or not good QCC predicted perturbatively 
remains for a considerable amount of time.
If so, then the perturbative treatment provides a useful
guide to our understanding of QCC in intrinsic decoherence dynamics.
If not, then our perturbative results make sense only for
extremely weak decoherence.  Dramatically, our computational studies 
strongly support our analytical perturbation results, even in the presence of
significant decoherence. For example, 
consider the case, where the parameters for the initial state are
the same as in
the previous case (therefore the initial state is also much delocalized), 
but the coupling potential is given by $V_{12}(Q,q)=Q^{2}\sin^{2}(q)$.
This coupling potential is highly nonlinear in $q$ but still quadratic in $Q$.
In accord with the second-order perturbation
results, such a coupling potential should still give rise to
good early-time QCC in the intrinsic decoherence dynamics of the first subsystem.
This is confirmed by the quantitative comparison between $S_{{\rm q}}(t)$ and $S_{c}(t)$
shown in Fig. \ref{nonlinear2}. More importantly,
Fig. \ref{nonlinear2} shows that outstanding
QCC remains even when both $S_{{\rm q}}(t)$ and $S_{c}(t)$ have increased
to close to their saturation value of unity.
Numerous other computational results (not shown) are consistent with the two cases shown here.

These results  
show the usefulness of the second-order
perturbation theory
in understanding QCC in intrinsic decoherence dynamics emanating from squeezed initial states.
Also of interest is intrinsic decoherence dynamics associated with
sufficiently localized initial states, which, in accord with the previous perturbation results,  
should display excellent early-time QCC for any coupling potential
$V_{12}(Q,q)$.  
We now 
computationally examine QCC at later times for localized states as initial conditions
and explain the results in terms of a simple classical theory of intrinsic decoherence dynamics.

\section{Localized Initial States}
\label{classical-theory}

\subsection{Simple Classical Approach}

Below we show that $S_{{\rm q}}(t)$ and $S_{c}(t)$ are often in
excellent agreement, over large time scales, for initially localized
states, significantly extending the perturbation theory result.
In doing so we compare the quantum $S_{{\rm q}}(t)$ with full
classical mechanics as well as with  a simple classical theory derived
in this section. The latter provides further insight into the origins
of increasing $S_{c}(t)$.

To derive the simplified classical result  we first use Liouville's theorem to
reexpress  $S_{c}(t)$ as
\begin{eqnarray}
 S_{c}(t)&=& 1-2\pi\hbar\int 
 \int \rho_{c}(Q(t),P(t),q(t),p(t),t) \nonumber \\
 & & \times 
 \rho_{c}(Q(t),P(t),q',p',t)\ dq'\ dp'\  dQ(t)\ dP(t)\ dq(t)\ dp(t) \nonumber \\
 &=& 1-2\pi\hbar\int \int \rho_{c}(Q,P,q,p,0)
  \rho_{c}(Q(t),P(t),q',p',t)\ dq'\ dp'  \ dQ\ dP\ dq\ dp  \nonumber \\
  &=& 1-2\pi\hbar\int \int \rho_{c}(Q,P,q,p,0)
    \rho_{c}(Q'',P'',q'',p'',0)\ dq'\ dp'\ dQ\ dP\ dq\ dp,
    \label{seq1}
  \end{eqnarray}
where $(Q(t),P(t),q(t),p(t))$ is the phase space location of the trajectory 
emanating from $(Q,P,q,p)$ at $t=0$,
and $(Q'',P'',q'',p'')$ is the phase space location of the trajectory at time zero
if the classical trajectory is propagated backwards from $(Q(t),P(t),q',p')$.
Because the initial state $\rho_{c}(Q,P,q,p,0)$ 
is assumed highly localized in phase
space,  
$(Q'',P'',q'',p'')$ must be very close to $(Q,P,q,p)$ in order for 
the term $\rho_{c}(Q,P,q,p,0)
\rho_{c}(Q'',P'',q'',p'',0)$ in Eq. (\ref{seq1}) to be appreciable and thus
to contribute to $S_{c}(t)$. 
Hence a convenient approximation can be made: 
we assume that, at time $t$,
only those backward trajectories near $(Q(t),P(t),q(t),p(t))$ need be taken into account.
This means that we treat
$Q''-Q$, $P''-P$, $q''-q$, $p''-p$,
$\delta q' \equiv [q'-q(t)]$, and $\delta p'\equiv [p'-p(t)]$  as sufficiently small
such that
\begin{eqnarray}
Q''&\approx&Q+M_{13}(t)\delta q' + M_{14}(t)\delta p',  \nonumber \\
P''&\approx&P+M_{23}(t)\delta q' + M_{24}(t)\delta p',  \nonumber \\
q''&\approx &q+M_{33}(t)\delta q' + M_{34}(t)\delta p',  \nonumber\\
p''&\approx &p+ M_{43}(t)\delta q' + M_{44}(t)\delta p',
\label{seq2}
\end{eqnarray}
where $M_{ij}$ ($i,j=1,2,3,4$) is
the stability matrix associated with the backward trajectories emanating from
$(Q(t),P(t),q(t),p(t))$:
\begin{eqnarray}
M_{ij}=\frac{\partial (Q, P, q, p)}{\partial (Q(t), P(t), q(t), p(t))}.
\label{seq3}
\end{eqnarray}
Although this approximation should be less reliable 
for chaotic systems, we demonstrate below that, it is, nonetheless,
computationally
useful in both integrable and chaotic cases.

For simplicity we consider below a specific example
in which $\tilde{\rho}_{1}^{0}(Q,P)$ and $\tilde{\rho}^{0}_{2}(q,p)$ are
symmetric Gaussian states given by
Eq. (\ref{initialstate}) with $\sigma_{Q}=\sigma_{P}=$$\sigma_{q}=\sigma_{p}\equiv \sigma=\sqrt{\hbar/2}$.
Other distributions can be considered in an analogous fashion.
Substituting Eqs. (\ref{initialstate}) and (\ref{seq2}) into
Eq. (\ref{seq1}) and evaluating the integrals, we obtain
\begin{equation}
S_{c}(t)=1-\frac{1}{2}\left\langle \frac{\exp\left[
\frac{U^{2}X^{2}+V^{2}Y^{2}-2UVZ}{2\sigma^{2}(X^{2}Y^{2}-Z^{2})}\right]}{\sqrt{
X^{2}Y^{2}-Z^{2}}} \right\rangle_{\rho_{c}'},
\label{seq4}
\end{equation}
where
\begin{eqnarray}
X&=&M_{13}^{2}+M_{23}^{2}+M_{33}^{2}+M_{43}^{2}, \nonumber \\
Y&=& M_{14}^{2}+M_{24}^{2}+M_{34}^{2}+M_{44}^{2}, \nonumber \\
Z&=& M_{13}M_{14}+M_{23}M_{24}+M_{33}M_{34}+M_{43}M_{44}, \nonumber \\
U&=& (Q-Q_{0})M_{14}+
(P- P_{0})M_{24}+
(q- q_{0})M_{34}+
(p-p_{0})M_{44}, \nonumber \\
V&=& (Q- Q_{0})M_{13}+
(P- P_{0})M_{23}+
(q-q_{0})M_{33}+
(p- p_{0})M_{43}, 
\label{seq5}
\end{eqnarray}
and
\begin{eqnarray}
\rho_{c}'(Q,P,q,p)=4\pi^{2}\hbar^{2}[\tilde{\rho}_{1}^{0}(Q,P)]^{2}[\tilde{\rho}_{2}^{0}(q,p)]^{2}.
\end{eqnarray}
Equations (\ref{seq4}) and (\ref{seq5}) indicate that the classical dynamics of 
intrinsic decoherence
is closely related to the classical stability matrix elements averaged over a rescaled
initial distribution
function.  
This interesting connection 
provides insight into
a variety of interesting aspects of quantum intrinsic  decoherence 
dynamics \cite{furuya,miller,arul,fujisaki,prosen}.
For example, 
for chaotic dynamics in which classical trajectories are highly unstable
and therefore in which
$|M_{ij}|$ increases rapidly,
$S_{c}(t)$ should increase much faster than for the case of integrable
dynamics.  This observation can, 
with the assumption that there is fairly good QCC in intrinsic decoherence
dynamics, directly explain
previous results on quantum signatures of classical chaos
in the dynamics of quantum entanglement  
\cite{furuya}. Further, 
because Eqs. (\ref{seq4}) and (\ref{seq5}) are expressed in terms of classical
stability matrices,
 characteristics of the time
dependence of $S_{c}(t)$ and therefore of $S_{{\rm q}}(t)$ can be easily related to
the time and space fluctuations in the instability of classical trajectories.

\subsection{Computational Results: Localized Initial States}

\begin{figure}[ht]
\begin{center}
\epsfig{file=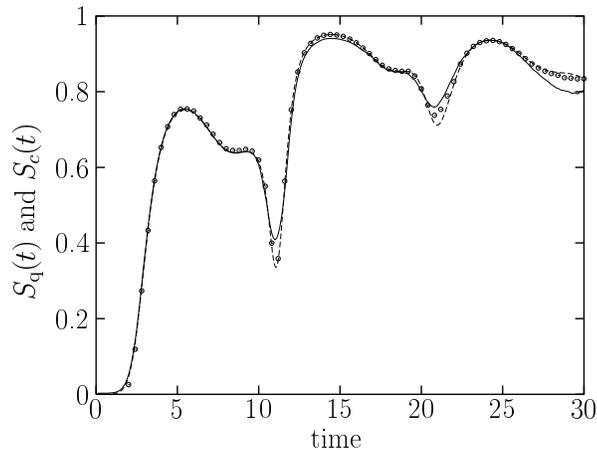,width=9.0cm}
\end{center}
\caption{ A comparison between
$S_{{\rm q}}(t)$ (dashed line) and the approximate $S_{c}(t)$ (solid line) calculated
from Eq. (\ref{seq4})
for the quartic oscillator model in the case of integrable dynamics ($\alpha=0.03,
\beta=0.01$).  The initial state is given by Eq. (\ref{initialstate}), with
$\sigma_{P}=\sigma_{Q}=\sigma_{p}=\sigma_{q}=\sqrt{\hbar/2}$,
$\hbar=0.005$,
$Q_{0}=0.4$, $P_{0}=0.5$, $q_{0}=0.6$,
and $H(Q_{0}, P_{0}, q_{0}, p_{0})=0.24$.
Full classical results based upon Eq. (\ref{seq1}) are represented by discrete
circular points.
All variables
are in dimensionless units.
}
\label{Fig-inte}
\end{figure}

\begin{figure}[ht]
\begin{center}
\epsfig{file=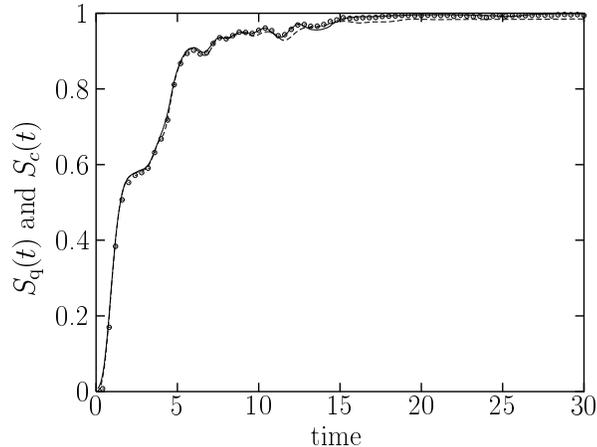,width=9.0cm}
\end{center}
\caption{Same as in Fig. \ref{Fig-inte} except for strongly chaotic dynamics ($\alpha=1.0, \beta=0.01$).}
\label{Fig-chaotic}
\end{figure}

Consider then QCC over large time scales for localized initial states
for both integrable and chaotic cases. To do so
we examine 
the quartic oscillator model as well as results where
the coupling potential is replaced by
the nonlinear potential $V_{12}(Q,q)=0.5Q^{2}q^{2}+Q^{4}q^{2}$.
The initial states are chosen to be localized initial states,  and both
the full classical dynamics and the 
approximate time dependence of $S_{c}(t)$ 
in Eq. (\ref{seq4})  are compared to the quantum result.
Specifically,  we realize the ensemble average in
Eq. (\ref{seq4}) by Monte-Carlo simulations, using
only $2\times 10^{4}$ sampling
classical trajectories from which the stability matrix elements $M_{ij}$ are evaluated.
The initial Gaussian states are chosen symmetric
with $\sigma=\sqrt{\hbar/2}=0.05$ and are sufficiently localized so that
Eq. (\ref{seq2}) should be a valid approximation. 
Full classical results for $S_{c}(t)$ (represented again by discrete circular points), which
are much more demanding computationally,
are also provided below.

Figures \ref{Fig-inte} and \ref{Fig-chaotic}
compare results for $S_{c}(t)$ obtained from 
Eq. (\ref{seq4}) and from exact classical results
with $S_{{\rm q}}(t)$ (dashed line)
for integrable and chaotic
dynamics in the quartic oscillator model,  respectively.  
A number of observations are in order.
First, it is clear that in both cases the approximate $S_{c}(t)$
are in excellent agreement with the full classical results, confirming
the utility of the simple model [Eq. (\ref{seq4})].
Second, both $S_{c}(t)$ and $S_{{\rm q}}(t)$ are seen, 
in the chaotic case, to relax faster towards $1.0$
than they do in the integrable case. 
Third,  the oscillation amplitudes of $S_{{\rm q}}(t)$
in the chaotic case are much smaller than that in the integrable case.
Hence, the fast relaxation and small-amplitude oscillations 
of $S_{{\rm q}}(t)$ shown in Fig. \ref{Fig-chaotic}
may be regarded as fingerprints of the underlying classical chaos.
Finally, and most importantly, the entire time dependence, including
oscillations in  Fig. \ref{Fig-inte}
and Fig. \ref{Fig-chaotic} of  $S_{{\rm q}}(t)$
are beautifully captured by both the exact and the approximate $S_{c}(t)$.
It should also be noted that 
the QCC time scale shown in Fig. \ref{Fig-chaotic}
is appreciably longer than is the QCC break time $t_{b}\sim 5.0$ for the same $\hbar$, 
obtained
by quantitatively comparing the structure of the
classical and quantum distribution functions
\cite{gongthesis}. 
This can be understood by the fact 
that $S_{c}(t)$ [or $S_{{\rm q}}(t)$]
describes the reduced distribution functions $\tilde{\rho}_{c}(Q,P,t)$ [or $\tilde{\rho}_{W}(Q,P,t)$]
which is insensitive to the fine structure of  $\rho_{c}(Q,P,q,p,t)$
[or $\rho_{W}(Q,P,q,p,t)$].

Calculations for many other initial states confirm 
that the QCC results shown in 
Figs. \ref{Fig-inte} and \ref{Fig-chaotic} are typical,
indicating that (a) QCC is essentially exact over large time scales and (b)
the simple classical 
theory of intrinsic decoherence dynamics introduced above
provides a useful approximation to
the exact results. 

\begin{figure}[ht]
\begin{center}
\epsfig{file=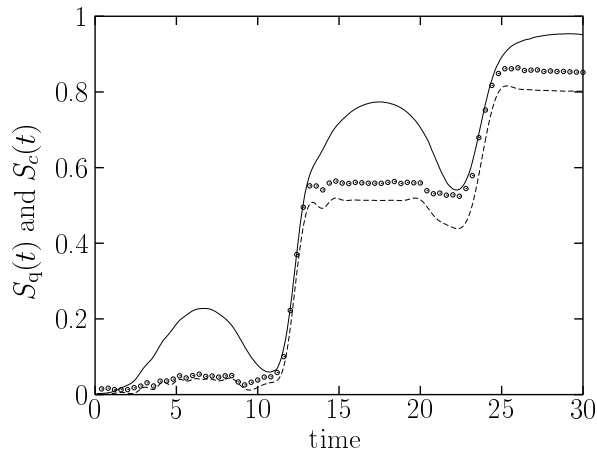,width=9.0cm}
\end{center}
\caption{Same as in Fig. \ref{Fig-inte} except for
strongly chaotic dynamics ($\alpha=1.0
, \beta=0.01$), and for a special initial Gaussian state, with
$Q_{0}=P_{0}=0$,  $q_{0}=0.6$ and  $H(Q_{0}, P_{0}, q_{0}, p_{0})=0.24$.}
\label{Fig-chan}
\end{figure}

Figure \ref{Fig-chan} shows one case, however,
where the approximate $S_{c}(t)$ and exact classical or quantum
results differ quantitatively. Here
the system is still the quartic oscillator model with $\alpha=1.0$, $\beta=0.01$, but
with an initial state of special type. In particular, both the initial average position $Q_{0}$
and the initial average momentum $P_{0}$ are set to zero.  Initial states of this type
are called channel states \cite{channel}, and
effectively give rise to very weak coupling
between the two subsystems over a considerably large time scale.  
Indeed, since at short times $df/d\overline{Q} \approx \Delta
f(\overline{Q})/\Delta Q\approx 0$ for $\overline{Q}\approx Q_{0}=0$, 
one obtains from 
Eqs. (\ref{cfg}) and (\ref{qfg}) that 
$1/\tau_{c,2}^{2}=1/\tau_{2,{\rm q}}^{2}\approx 0$.
Therefore the early-time intrinsic decoherence rate  should be small, as
seen in Fig. \ref{Fig-chan}, although the underlying classical dynamics is strongly chaotic.
For this reason one expects that 
the dynamical behavior of $S_{c}(t)$ and $S_{{\rm q}}(t)$ should 
differ from previous cases.
As shown in  Fig. \ref{Fig-chan}, in this case both 
$S_{{\rm q}}(t)$ and $S_{c}(t)$ increase in a step-wise fashion, distinctly
different from that in Figs. \ref{Fig-inte}
and \ref{Fig-chaotic}. Agreement between them remains excellent.
However, the approximate 
$S_{c}(t)$ misses some of the important structure. 

\begin{figure}[ht]
\begin{center}
\epsfig{file=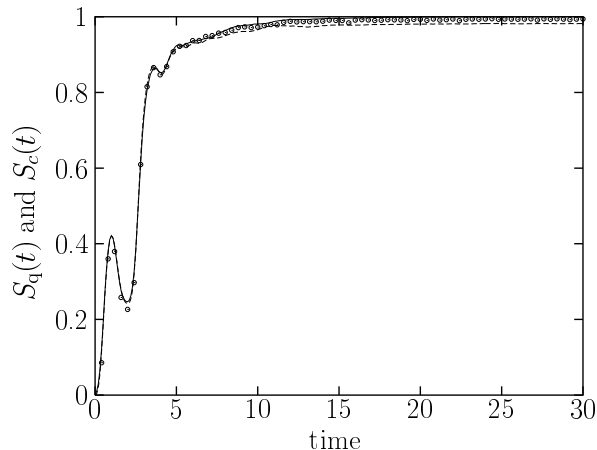,width=9.0cm}
\end{center}
\caption{Same as in Fig. \ref{Fig-inte} except for
a modified
quartic oscillator model in which
$V_{12}(Q,q)=0.5Q^{2}q^{2}
+Q^{4}q^{2}$.
}
\label{Fig-nonlinear}
\end{figure}

In Fig. \ref{Fig-nonlinear} we show the QCC result for a simple variant of the quartic oscillator model,
i.e., $V_{12}(Q,q)=0.5 Q^{2}q^{2}+Q^{4}q^{2}$, a coupling potential that is
neither linear nor quadratic. 
As seen in Fig.  \ref{Fig-nonlinear},
even with such nonlinear coupling 
$S_{c}(t)$ and $S_{{\rm q}}(t)$ are in excellent agreement
over large time scales.  This emphasizes the fact
that the good QCC results 
observed in the quartic oscillator model are not due to the fact
that the coupling potential
therein is quadratic.
Hence,
we conclude that for initially localized states,
our simple classical theory of
intrinsic decoherence dynamics [see Eq. (\ref{seq4})] is generally useful
in describing intrinsic decoherence dynamics in smooth Hamiltonian systems.

\section{discussion and summary}
\label{summary}

Quantum entanglement between individual subsystems
has no classical analog.
Nevertheless,  as shown in this work, the quantum dynamics of quantum entanglement, 
as manifest in the quantum dynamics of intrinsic
decoherence, does have a classical analog in 
classical Liouville dynamics describing classical correlations between
classical subensembles.
Hence it is useful to isolate the conditions under which 
there is good QCC in intrinsic decoherence dynamics. This is done analytically, for early-time dynamics
and for weak decoherence,
by a second-order
perturbative theory. Interestingly, as demonstrated by our computational studies
in Sec. \ref{sample} and Sec. \ref{classical-theory}, the physical picture of QCC
afforded by the perturbative treatment can be still very useful even when the time scale under
investigation is relatively long and the degree of intrinsic decoherence is significant.
In particular,
under the circumstances where there is good early-time QCC, 
classical Liouville dynamics can provide a simple means of understanding
different aspects of intrinsic decoherence dynamics, for relatively large time scales and
for both integrable and chaotic dynamics.
Further,
we have derived an approximate but very simple
classical theory of linear entropy production of
intrinsic decoherence dynamics associated with localized initial states, 
and shown that the rate of entropy production is closely
related to the stability properties of classical trajectories.

Clearly, the linear entropy is just one of many possible representation-independent
measures of intrinsic decoherence, 
and $S_{{\rm q}}(t)\approx S_{c}(t)$ does not mean that
the quantum dynamics is equivalent to the corresponding classical Liouville dynamics.
For example, if the saturation value of the linear entropy
in the long time limit is of particular interest, then the measures 
$1/[1-S_{c}(t)]$ and $1/[1-S_{{\rm q}}(t)]$ (which gives the number of orthogonal states that are incoherently
populated) should be more useful in describing QCC.
Indeed,
our results in Figs. \ref{Fig-chaotic} and \ref{Fig-nonlinear} suggest that
as time increases one has  $1/[1-S_{c}(t)]>>1/[1-S_{{\rm q}}(t)]$.
This is consistent with our previous observation \cite{gongprl} that, 
decoherence can dramatically improve QCC, but 
even strong decoherence does not necessarily suffice to ensure that quantum entropy
production is the same as classical entropy production.

\begin{figure}[ht]
\begin{center}
\epsfig{file=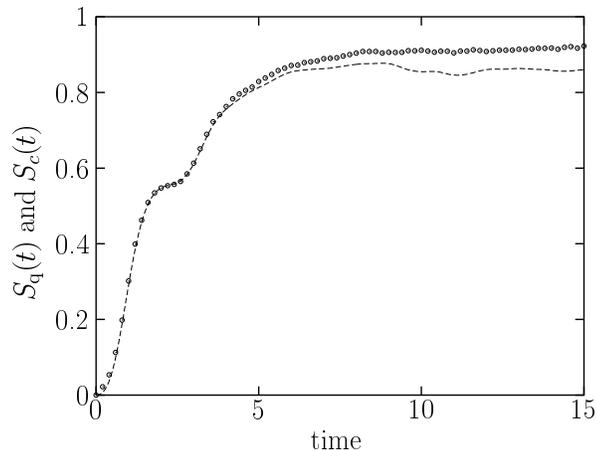,width=9.0cm}
\end{center}
\caption{A comparison between $S_{{\rm q}}(t)$ (dashed line) and $S_{c}(t)$ (discrete
circular points)
for strongly chaotic dynamics
of the quartic oscillaor model ($\alpha=1.0
, \beta=0.01$) and for $\hbar=0.05$.
The initial state is given by Eq. (\ref{initialstate}), with
$\sigma_{P}=\sigma_{Q}=\sigma_{p}=\sigma_{q}=\sqrt{\hbar/2}$,
$Q_{0}=0.4$, $P_{0}=0.5$, $q_{0}=0.6$,
and $H(Q_{0}, P_{0}, q_{0}, p_{0})=0.24$.
All variables
are in dimensionless units.
}
\label{bighbar}
\end{figure}

It should also be pointed out that
the model quantum systems studied in this paper are still far from 
the semiclassical regime.
This is indicated, in the chaotic case of the quartic oscillator 
model for example, by the fact that the QCC break time 
is relatively short compared to the time scale that we examined. Correspondence
will worsen quantitatively with increasing $\hbar$, although,
as discussed above, $\hbar$ 
is far from the only factor influencing the quality of the QCC.
However, we note that for localized initial states the qualitative features of the time dependence of
classical and quantum linear entropy may remain similar to one another
with much larger effective Planck
constants.  For example, Fig. \ref{bighbar} displays fairly
good QCC between $S_{{\rm q}}(t)$ and $S_{c}(t)$,
in the chaotic case of the quartic oscillator model, with $\hbar=0.05$ and
with an initial
symmetric Gaussian state.

A number of interesting extensions of this work are under
consideration. First, it seems straightforward but necessary to consider cases in which the
coupling potential depends upon both position and
momentum. Second, we
propose to further investigate the role of the dynamics of 
the subsystems in addition to that of the coupling potential (e.g.,
the dynamics in coupled Morse oscillator systems). Third,
it is interesting to study QCC
in intrinsic decoherence dynamics in terms of the decay of off-diagonal
density matrix elements. Such studies are ongoing, with preliminary studies \cite{han} 
indicating that comparing the time dependence of  off-diagonal
density matrix elements to its direct classical analog \cite{gongprl} will
provide deeper insights into QCC in
intrinsic decoherence dynamics.

To summarize, we have shown that classical dynamics can be very useful in describing
intrinsic decoherence dynamics in smooth Hamiltonian systems.  In particular, we have identified
conditions under which excellent
quantum-classical correspondence 
in the early-time dynamics of intrinsic decoherence is
possible via a second-order perturbative treatment,
have presented a simple classical theory 
of intrinsic decoherence dynamics emanating from localized initial states, and
have provided supporting computational results.
The hope is that by extending this study to high-dimensional Hamiltonian systems,  
we may use purely classical approaches to describe (at least qualitatively)
the dynamics of quantum entanglement or
intrinsic decoherence in polyatomic molecular systems.

{\bf Acknowledgments:} This work was supported by the U.S. Office of
Naval Research
and by the Natural Sciences and Engineering
Research Council of Canada.


\begin{thebibliography}{100}
\bibitem{zurekreview} W.H. Zurek, preprint quant-ph/0105127.
\bibitem{batista} V.S. Batista and P. Brumer, \prl{\bf 89}, 143201 (2002).
\bibitem{qcbook} M.A. Nielsen and I.L. Chuang, {\it Quantum Computation
and Quantum Information} (Cambridge University Press, Cambridge, 2000).
\bibitem{brumerreview}
M. Shapiro and P. Brumer, Adv. Atom. Mol. and Opt.
Phys., {\bf 42}, 287 (2000).
\bibitem{ricebook}
S.A. Rice and M. Zhao, {\it Optical Control of Molecular Dynamics} (John
Wiley, New York, 2000).
\bibitem{brumerbook}M. Shapiro and P. Brumer, {\it Principles of
the Quantum Control of Molecular Processes} (John Wiley, New York, 2003).
\bibitem{wilkie} J. Wilkie and P. Brumer, Phys. Rev. A {\bf 55}, 27 (1997);
Phys. Rev. A {\bf 55}, 43 (1997).
\bibitem{gongprl} J. Gong and P. Brumer, Phys. Rev. Lett. {\bf 90},
050402 (2003);
J. Gong and P. Brumer, quant-ph/0212106.
\bibitem{bmiller}
H. Wang, M. Thoss, K.L. Sorge, R.X. Gim\'{e}nez, and
W.H. Miller, \jcp{\bf 114}, 2562 (2001);
F. Grossmann, \jcp{\bf 103}, 3696 (1995);
A.M.O. de Almeida, preprint quant-ph/0208094.
\bibitem{linear-entropy}
P.C. Lichtner and J.J. Griffin, \prl{\bf 37}, 1521 (1976);
W.H. Zurek, S. Habib and J.P. Paz, \prl{\bf  70}, 1187 (1993);
X-P. Jiang and P. Brumer, Chem. Phys. Lett. {\bf 208}, 179 (1993);
a. Isar, A. Sandulescu, and W. Scheid,
\pre{\bf 60}, 6371 (1999);
G. Manfredi and M. R. Feix, \pre{\bf 62}, 4665 (2000);
J.N. Bandyopadhyay and A. Lakshminarayan, \prl{\bf 89}, 060402 (2002).
\bibitem{moyal} J.E. Moyal, Proc. Cambridge Phil. Soc. {\bf 45}, 99 (1949).
\bibitem{kim} J.I. Kim, M.C. Nemes, A.F.R. Toledo Piza, and H.E. Borges, \prl{\bf
77}, 207
(1996).
\bibitem{duan} L.M. Duan and G.C. Guo, \pra{\bf 56},
4466 (1997).
\bibitem{bacon}D. Bacon, D.A. Lidar, and K.B. Whaley, \pra{\bf 60},
1944 (1999).
\bibitem{eckhardt} B. Eckhardt, G. Hose, and E. Pollak, \pra{\bf 39}, 3776
(1989).
\bibitem{gongpre} J. Gong and P. Brumer, \pre{\bf 60}, 1643 (1999).
\bibitem{channel}
M.S. Santhanam, V.B. Sheorey, and A. Lakshminarayan, \pre{\bf 57}, 345 (1998).
\bibitem{furuya}K. Furuya, M.C. Nemes, and G.Q. Pellegrino, \prl{\bf 80}, 5524 (1998).
\bibitem{miller}P.A. Miller and S. Sarkar, \pre{\bf 60}, 1542 (1999).
\bibitem{arul} A. Lakshminarayan, \pre{\bf 64}, 036207 (2001).
\bibitem{fujisaki} A. Tanaka, H. Fujisaki, and T. Miyadera, \pre{\bf 66}, 045201 (2002). 
\bibitem{prosen}M. Znidaric and T. Prosen, J. Phys. A {\bf 36}, 2463 (2003).
\bibitem{gongthesis} J. Gong, Ph.D thesis (unpublished), University of Toronto, 2001.
\bibitem{han} H. Han, J. Gong, and P. Brumer, to be published.
\end{thebibliography}
\end{document}